\documentclass[final]{elsarticle}

\usepackage{lineno,hyperref}
\modulolinenumbers[5]

\journal{Journal of Contaminant Hydrology}

\usepackage{amsmath}
\usepackage{bm}
\usepackage{bbm}
\usepackage{amsfonts}
\usepackage{amssymb}
\usepackage[usenames,pdftex,dvipsnames,svgnames,table]{xcolor}
\usepackage{commath}
\usepackage{ulem}

\usepackage{xargs}                      
\usepackage{fullpage,lmodern}
\usepackage[inline]{showlabels,rotating}

\newcommand{\SH}{{\rm Sh}}
\newcommand{\PE}{{\rm Pe}}

\newcommand{\NU}{{\rm Nu}}
\newcommand{\DA}{{\rm Da_{I}}}
\newcommand{\DAD}{{\rm Da_{II}}}

\usepackage{tikz}
\usetikzlibrary{hobby,patterns}
\pgfdeclarepatternformonly{north east lines wide}%
   {\pgfqpoint{-1pt}{-1pt}}%
   {\pgfqpoint{6pt}{6pt}}%
   {\pgfqpoint{5pt}{5pt}}%
   {
     \pgfsetlinewidth{0.4pt}
     \pgfpathmoveto{\pgfqpoint{0pt}{0pt}}
     \pgfpathlineto{\pgfqpoint{5.1pt}{5.1pt}}
     \pgfusepath{stroke}
    }
\definecolor{applegreen}{rgb}{0.55, 0.71, 0.0}

\newcommand{\Ub}{\mathbf{U}}
\newcommand{\ub}{\mathbf{u}}

\newcommand{\xb}{\mathbf{x}}
\newcommand{\Xb}{\mathbf{X}}

\newcommand{\nb}{\mathbf{n}}

\newcommand{\wienerb}{\mathbf{W}}

\newcommand{\be}[1]{
\begin{equation}
\expandafter\label{eq:#1}
}
\expandafter \newcommand{\ee}{\expandafter\end{equation}}
\expandafter \newcommand{\eq}[1]{Eq.\expandafter(\ref{eq:#1})}

\expandafter \newcommand{\bfig}[1]{
\begin{figure}
\expandafter\label{fig:#1}
}
\expandafter \newcommand{\efig}{\end{equation}}
\expandafter \newcommand{\fig}[1]{Fig.~\ref{fig:#1}}

\newcommand{\f}{p}

\renewcommand{\u}{u}

\newcommand{\U}{U}
\newcommand{\C}{C}
\renewcommand{\c}{x}
\newcommand{\diff}{\mathcal{D}_{0}}
\newcommand{\disp}{\mathcal{D}_\textrm{eff}}

\renewcommand{\c}{c}
\renewcommand{\C}{C}
\newcommand{\poros}{\varepsilon}

\renewcommand{\div}{\nabla\cdot}
\newcommand{\divx}{\nabla_{\xb}\cdot}

\newcommand{\grad}{\nabla}

\newcommand{\spatavg}[1]{\left<#1\right>} 
\newcommand{\ensavg}[1]{\overline{#1}} 

\newcommand{\Die}[1]{$10^{{#1}}$}

\newcount\colveccount
\newcommand*\colvec[1]{
        \global\colveccount#1
        \begin{pmatrix}
        \colvecnext
}
\def\colvecnext#1{
        #1
        \global\advance\colveccount-1
        \ifnum\colveccount>0
                \\
                \expandafter\colvecnext
        \else
                \end{pmatrix}
        \fi
}

\newcount\rowveccount
\newcommand*\rowvec[1]{
        \global\rowveccount#1
        \begin{pmatrix}
        \rowvecnext
}
\def\rowvecnext#1{
        #1
        \global\advance\rowveccount-1
        \ifnum\rowveccount>0
                &
                \expandafter\rowvecnext
        \else
                \end{pmatrix}
        \fi
}

\usepackage[colorinlistoftodos,prependcaption,textsize=small]{todonotes}
\newcommandx{\mat}[2][1=]{\todo[inline,linecolor=red,backgroundcolor=blue!25,bordercolor=red,#1]{MATTEO: #2}}
\newcommandx{\ele}[2][1=]{\todo[inline,linecolor=blue,backgroundcolor=green!25,bordercolor=blue,#1]{ELEONORA: #2}}
\newcommandx{\gian}[2][1=]{\todo[inline,linecolor=OliveGreen,backgroundcolor=CarnationPink!25,bordercolor=OliveGreen,#1]{GIANLUCA: #2}}

\begin{document}
\graphicspath{{Data/}{./}}

\begin{frontmatter}

\title{A robust upscaling of the effective particle deposition rate in porous media
}


\author[TAU]{Gianluca Boccardo}

\author[polito]{Eleonora Crevacore}
\author[polito]{Rajandrea Sethi}

\author[warwick]{Matteo Icardi\corref{mycorrespondingauthor}
}
\ead{m.icardi@warwick.ac.uk}
\cortext[mycorrespondingauthor]{Corresponding author}

\address[TAU]{School of Mechanical Engineering, Tel Aviv University, Tel Aviv, 69978, Israel}
\address[polito]{DISMA, Politecnico di Torino, C.so Duca degli Abruzzi 24, Torino, Italy}
\address[diati]{DIATI, Politecnico di Torino, C.so Duca degli Abruzzi 24, Torino, Italy}
\address[warwick]{Mathematics Institute, University of Warwick, CV4 7AL, Coventry, UK}

\begin{abstract}
In the upscaling from pore- to continuum (Darcy) scale, reaction and deposition phenomena at the solid-liquid interface of a porous medium have to be represented by macroscopic reaction source terms.
The effective rates can be computed, in the case of periodic media, from three-dimensional microscopic simulations of the periodic cell.
Several computational and semi-analytical models have been studied in the field of colloid filtration to describe this problem.
They often rely on effective deposition rates defined by simple linear reaction ODEs, neglecting the advection-diffusion interplay, and assuming slow reactions (or large P\'eclet numbers).
Therefore, when these rates are inserted into general macroscopic transport equations, they can lead to conceptual inconsistencies and, therefore, often qualitatively wrong results.
In this work, we study the upscaling of Brownian deposition on face-centred cubic (FCC) spherical arrangements using a linear effective reaction rate, defined by volume averaging,  and a macroscopic advection-diffusion-reaction equation. The case of partial deposition, defined by an attachment probability, is studied and the limit of ideal deposition is retrieved as a particular case.
We make use of a particularly convenient computational setup that allows the direct computation of the asymptotic stationary value of effective rates. This allows to drastically reduce the computational domain down to the scale of the single repeating periodic unit: the savings are ever more noticeable in the case of higher P\'eclet numbers, when larger physical times are needed to reach the asymptotic regime, and thus, equivalently, a much larger computational domain and simulation time would be needed in a traditional setup.
We show how this new definition of deposition rate is more robust and extendable to the whole range of P\'eclet numbers; it also is consistent with the classical heat and mass transfer literature.
\end{abstract}


\end{frontmatter}

\section{Introduction}\label{sec:intro}
Particle transport and deposition are fundamental phenomena behind several natural and engineered processes.
One of the many examples is the fate of pollutants in groundwater systems, an emerging environmental issue countered with interventions based on the injection of nanoscopic zero-valent iron particles, to cite a particular successful application~\citep{Velimirovic2016713,Krol2013,VecchiaLunaSethi}.
More in general, the study of solute deposition is of central importance in filtration processes to enhance air and water quality \citep{tiraferri2011transport}, chromatographic systems, catalytic cells and packed bed reactors~\citep{kolakaluri2015filtration,Bensaid2010357,dixon2001cfd}, enhanced oil recovery techniques~\citep{Shi2013}, and even drug delivery studies~\citep{Pankhurst2003,Gordon2014}: all these processes rely on a detailed understanding of how transported solutes/particles flow through a porous matrix and interact with it.
Thus in this section we will give a brief overview of the theoretical framework typically used in the study of mass transport and particle deposition in porous media, and, in particular, the classical \textit{colloid filtration theory}.
Secondly, we will touch upon the issues that affect the correlations commonly used to predict deposition efficiency and some inconsistencies in the process of upscaling the heterogeneous reaction at the pore-scale to a homogeneous reaction term in a macroscopic transport equation.
This will form the groundwork for the following sections, where a robust and mathematically sound methodology for the calculation and upscaling of deposition efficiency will be proposed, constituting the main contribution of this work; results from micro-scale fluid dynamic simulations are proposed along with it in the last section.

\subsection{Mass transfer and particle deposition}
In the dimensional analysis of mass transfer phenomena, the most used dimensionless quantity is the Sherwood number, which describes the ratio between convective mass transfer and diffusive transport, and is the analogous of the Nusselt number used in heat transfer.
It can be defined as:
\begin{equation}\label{eq:Sherwood}
\SH=\frac{K L}{\mathcal{D}}
\end{equation}
where  $L$ is a characteristic length (m), and $\mathcal{D}$ is the molecular diffusion coefficient (m$^2$ s$^{-1}$).
K (m s$^{-1}$) is the same coefficient commonly used in the mass transfer equation:
\begin{equation}\label{eq:massTransfEq}
I= K\Delta C \mathcal{S}
\end{equation}
where $I$ is the molar flux (mol s$^{-1}$), $\mathcal{S}$ the effective mass transfer surface (m$^2$), and $\Delta C$ the concentration driving force (mol m$^{-3}$).
In most mass transfer applications in porous media, the characteristic length is taken to be equal to the effective grain diameter $d_\textrm{g}$.
As such, the mass transfer is then characterised as
$
\SH=I d_\textrm{g}/\left(\mathcal{D}\Delta C \mathcal{S}\right)
$.

In this work we will consider the case of solute deposition (or, equivalently, filtration).
A wide bibliography is available on this topic, and the approach most commonly employed in order to determine a single parameter describing the filter effectiveness from its features and the operating conditions under investigation, is to define a \textit{collector efficiency} $\eta$~\citep{yao1968,yao1971,logan1995}.
This \textit{total} efficiency coefficient is expressed as the contribution of two terms:
%
$
\eta=\alpha \eta_0
$, 
%
where $\alpha$ is the attachment efficiency, describing the probability of a particle colliding with the solid grain being adsorbed, with $0 < \alpha < 1$ depending on the specific physico-chemical features of the system.
The second term $\eta_{0}$ describes the migration of the particles from the bulk of the fluid to the surface of the grains, which is usually thought of as a contribution of different mechanisms, namely \underline{B}rownian diffusion, sterical \underline{i}nterception and inertial (and \underline{g}ravitational) effects. Furthermore, it is typically assumed that these contributions are additive~\citep{yao1971,PrieveRuckenstein1974}:
\begin{equation}\label{eq:EtaZeroAdditive} \nonumber
\eta_0=\eta_B + \eta_I + \eta_G
\end{equation}

Many efforts in colloid filtration theory have been devoted to the precise quantification of the this efficiency $\eta$ in specific micro-scale models, and its expression as a function of macro-scale parameters.
The earliest studies, by \citet{Levich}, dealing with diffusion on a solid sphere immersed in an infinite flow field moving with creeping flow, resulted in the evaluation of the molar flux towards the grains as:
\begin{equation}\label{eq:LevichFluxI}
I=7.98 C_{\infty} \mathcal{D}_0^{\frac{2}{3}}U^{\frac{1}{3}}a^{\frac{4}{3}}
\end{equation}
where $a$ is the solid grain radius (m), $C_{\infty}$ is the upstream solute concentration, and $U$ is the fluid upstream approach velocity (m s$^{-1}$). 
Defining the deposition efficiency as the ratio between the molar flux to the grains and the advective molar flux leads to~\footnote{It has to be noted that the approximated numerical coefficients in this and preceding equation are not coming from empirical estimations, but result from the analytical evaluation of volume integrals in Levich's axysymmetric single sphere model development. For the breakdown of all the steps, refer to~\citet{Levich}(Section 14, \textit{``Diffusion to a free-falling solid particle''}).}:
\begin{equation}\label{eq:EtaDefinitionADV}
\eta_B=\frac{I}{\pi d_\textrm{g}^2 U C_{\infty}} = \eta_B= 4.04 Pe^{-\frac{2}{3}}
\end{equation}

One issue with this model, aside from the clear impossibility of representing a randomly packed bed as a collection of isolate spheres,
lies in the particular boundary condition employed by Levich of solute concentration being equal to upstream concentration on the surface of the grain at the impinging point \citep{Levich}.
This comes from the assumption of advection being dominant over diffusion, which limits the usefulness of this expression (and others, built on this same simplification) to $\PE\gtrsim70$.

In order to account for the packed bed topology and especially for the effect that neighbouring grains have on the filtration efficiency of a single collector, Pfeffer and Happel in their seminal works \citep{pfeffer1964,pfeffer1964analytical}, obtained the following relation:
\begin{equation}\label{eq:PfefferSherwood}\nonumber
\SH=As^{\frac{1}{3}}\PE^{\frac{1}{3}}
\end{equation}
where $As$ is a porosity-dependent parameter equal to 
$
As=\frac{2(1-\gamma^5)}{2-3\gamma+3\gamma^5-2\gamma^6}
$, and $\gamma=(1-\poros^{1/3})$.
%
Considering the $As$ parameter and putting the last few expressions together, an expression for $\eta$ similar to Levich's relation can be obtained, i.e.: 

\begin{equation}\label{eq:EtaHappel}
\eta_B=4As^{\frac{1}{3}}\PE^{-\frac{2}{3}} \, .
\end{equation}

A great deal of work has been done over the years, based on the colloid filtration theory, to refine the understanding of solute fate and transport by improving these single collector models \citep{rajagopalantien1976,tufenkji2004,LongHilpert2009,johnson2013upscaling}, and then building a connection between the single collector efficiency $\eta$ calculated at the microscopic scale, and an upscaled reaction rate employable in a macroscopic transport equation, $K_\textrm{d}$.

In the next section, we will detail some of the issues with these studies, especially with regards to the assumptions considered in the derivation of the micro-scale models, and their impact on a successful upscaling.
For the sake of clarity, we will limit the following exposition to the case of Brownian deposition, but it has to be noted that a considerable amount of work has been done for the derivation of constitutive equations which took into account the effects of particle inertia, gravity, Van der Waals interactions and other physical interaction phenomena: the reader is referred to the extensive literature on the wider topic of colloid deposition.

\subsection{Upscaling particle deposition: the role of $\eta$ and $K_\textrm{d}$}
As it has been mentioned, the most widely used approach in the colloid deposition literature has been to study in detail simplified models representing a single collector, followed by a heuristic step providing the link between solute transport in the vicinity of one collector to the evolution of the phenomenon at the macro-scale.
The construction of the simplified model itself is of critical importance in order to avoid gross misrepresentations of the structure of the porous medium under consideration.
In the preceding section we have mentioned that the early models of the colloid filtration theory described the porous matrix as an assemblage of isolated spheres~\citep{Levich,yao1971}; this model was then superseded by Happel's sphere in cell model which inherently carries the information about the packing porosity and takes into account the effect of neighbouring grains on the transport around one collector.
In turn, a relatively modern description improved on Happel's model by substituting the single sphere with two touching hemispheres~\citep{Ma2009}: this seemingly simple change does correct for the glaring missing piece in Happel's models, via the introduction of contact points between different collectors, which has a noticeable influence on the evolution of the dynamics of solute deposition and, in a perhaps more recognisable problem from an engineering standpoint, heat transfer~\citep{WehingerKraume2017}.

What these models, though provenly effective, fail to grasp is the effect of grain surface heterogeneity and especially grain polidispersity on solute deposition; efforts to remedy this shortcoming are very recent~\citep{RasmusonPazmino2017}.
Nonetheless, the main issue which affects this current approach, common to the state of the art at large, is the link between the values of $\eta$ obtained by these (albeit very precise) representations, and a ``corresponding'' value of the reaction rate of the macroscopic removal/sink term, $K_\textrm{d}$.
These are all based on two fundamental quantities: the upstream concentration $C_\infty$ equal to the solute concentration far from the collector influence, and an outgoing or downstream concentration $C < C_\infty$.
The common approach is to write an acceptable relation expressing both $K_\textrm{d}$ and $\eta$ as a function of the ratio $C/C_\infty$, then equating the two resulting relations.
The first hypothesis is to assume steady-state transport, neglecting the transient term and thus considering a direct relationship between advection and particle removal (seen as a first-order process), leading to:
\begin{equation}\label{eq:A-F}
U\dfrac{d C}{d x}=K_\textrm{d}C \quad \Longrightarrow \quad \log \left( \dfrac{C}{C_\infty}\right)=-\dfrac{K_\textrm{d}}{U}L
\end{equation}
in an infinitesimal distance $dx$, where $C$ is an integral over a plane orthogonal to the main direction of the fluid moving with Darcy velocity $U$.

The following step requires an expression of the particle deposition efficiency $\eta$, depending on the concentration ratio $C/C_\infty$; a few different approaches exist.
The idea behind most studies, starting from the earliest works in the colloid filtration theory~\citep{yao1971}, is to consider the porous matrix as a pseudo-homogeneous medium, and similarly to Eq.~\ref{eq:A-F}, write an infinitesimal balance equation.
Then, neglecting again both the transient and the diffusion term, this results in the classical filter equation:
\begin{equation}\label{eq:FilterEq}
\frac{d C}{d x}= -\frac{3}{2} \frac{1-\poros}{\poros}\frac{1}{d_\textrm{g}}\alpha \eta C \quad
\Longrightarrow\quad
\log \left( \dfrac{C}{C_\infty}\right)=-\frac{3}{2}\frac{1-\poros}{\poros}\frac{L}{d_\textrm{g}}\alpha \eta \,.
\end{equation}
An in-depth, step-by-step explanation of this procedure is given by \citet{logan1995}.
An alternative simpler approach is to abandon the infinitesimal balance description and to construct the porous medium as a corrected sequence of ``single collector models'' each characterised by the general efficiency $\eta$~\citep{molnar2015,johnson2013upscaling,NelsonGinn2011}.
Following this conceptualization, it is possible to write:
\begin{equation}\label{eq:Eta-Ratio-HILPERT}
\log \left( \dfrac{C}{C_\infty}\right)=N_\textrm{C} \log(1-\eta) \, ,
\end{equation}
where $N_\textrm{C}$ is the number of collectors considered.
The integral expression in Eq.~\ref{eq:A-F} can then be equated with Eq.~\ref{eq:Eta-Ratio-HILPERT} and, depending on the geometrical model considered, and thus on the different relation between the ration $N_\textrm{C} / L$ and $\poros$, this leads to various expressions relating $K_\textrm{d}$ and $\eta$.
Given this background, it is now clear how the number of assumptions taken in the development of this upscaling heuristic can weigh on the final applicability of any proposed model, for a number of reasons.

\subsection{Issues with the present upscaling approach}

First, as it has been noted in the exposition just given, in the usual definition of $\eta$ (as in Eq.~\ref{eq:EtaDefinitionADV}), it is customary to neglect the diffusive contribution to the flux of particles towards the grain surface; this has also been already noticed by several authors \citep{messina2015, Crevacore2016271}, and makes the resulting model valid only in a certain range of operating conditions (i.e.: low P\'eclet numbers).

Another problem arising very strongly in the case of very low P\'eclet numbers, or imperfect mixing, lies in the evaluation of the actual concentration driving force: while it is usually (in the case of perfect-sink boundary conditions, at least) understood that $\Delta C = C_{\infty}$, this is in fact not accurate when the concentration gradient across the control pore-scale volume used as the basis for the upscaling procedure is not constant or uniform.

Another simplification, which is always inherently considered, is to neglect the existence of a pre-asymptotic (non constant) and asymptotic reaction scenario.
Similarly to the velocity profile, also the concentration profile, undergoing advection, diffusion, and reaction will converge to a self-similar solution and, therefore, a stationary value of deposition (reaction) efficiency can only be found after a (possibly long) pre-asymptotic regime.
This has been recently shown by \citet{messina2016}, without however reaching an explicit computation of the asymptotic value.


The two most important issues are however related to the calculation of deposition efficiency at the micro-scale, and its relation to the macroscopic reaction rate.
The former is often considered proportional to the deposition efficiency, while, from a simple infinitesimal mass balance, $K_\textrm{d}$ is clearly proportional to the logarithm of $1-\eta$. The two are equivalent only in the limit of efficiencies close to zero (slow reaction typically linked to high P\'eclet number).
This has been recognised, for example by~\citet{johnson2013upscaling}, but in many works the wrong upscaled term is still being used.

More importantly, a significant mistake is often committed in the derivation of a continuum equation, by wrongly applying an infinitesimal mass balance to an arbitrary size.
This makes the closure strongly dependent on the length scale considered.
This can be seen both when the elementary volume is relatively large, or when P\'eclet numbers are low.
Aside from the conceptual issues, there are clear numerical problems when dealing with cases with low P\'eclet numbers, which results in the particle concentration decaying to zero very fast and indeed on spatial scales smaller than even the geometrical system considered.
This, in turn, results in glaring errors in the estimation of $\eta$, as the working assumption is that the dimension of the \textit{active} filter volume in the x-axis is $L$\footnote{Where $L$ is the total length of the computational domain chosen.}, allowing to calculate the outlet concentration in that point and estimate deposition efficiency on that partial control volume.
However, when the deposition efficiency is unitary and thus the downstream concentration $C\rvert_{x=L}=0$, the practical implication is that the \textit{active} filter area is smaller than the chosen computational volume, making any estimation based on averages over the full domain incorrect, and an effective upscaling impossible.
Moreover, in cases such as the ones explored in these work, it is not possible to further shrink the integration area, as it would result in passing the lower threshold for the \textit{geometrical} representative elementary volume.

As a last point, the role of the attachment efficiency $\alpha$ is always taken into account only as a linear correction (scaling) of the overall reaction rate.
In  this work we show how a non-unitary attachment efficiency (i.e.: unfavourable physico-chemical conditions) has a more complicated non-linear effect on deposition rate.

As it can be seen, most of these inconsistencies clearly appear when the deposition is dominant (small P\'eclet numbers) and have been already recognised in the literature, but notwithstanding the many works dealing with a precise calculation of the deposition efficiency $\eta$ in a single collector, the issue of a robust upscaling still has not been fully explored.

We propose to solve these issues with the introduction of an effective reaction rate obtained via a simple, though rigorous, upscaling based on volume averaging. This has the benefit of being agnostic with respect to the specific definition of the single collector, allowing for an easier calculation and extension to a wide variety of different porous media structures, without the added encumbrance of being limited to a single simplified model for their description.
Particular care will be given to respect some minimal consistency requirements of a proper upscaling procedure.
First, consistent physical assumptions at both (micro/macro) scales have to be employed.
Then, the final upscaled parameters should be \textit{scale independent}. This means that, once it is assumed (or verified) that the micro-scale geometry is a representative elementary volume\footnote{For periodic geometries, a geometric REV corresponds to the single repeating elementary module.}, the upscaled parameters, appropriately normalised, should not depend on the size of the volume considered.

The remainder of the paper is organised as follows: the next section describes the physical assumptions and equations valid at the pore-scale. On Sec.~\ref{sec:macro} the steps of this process and the definition of an effective reaction rate are detailed.
This will be followed, in Sec.~\ref{sec:numerical}, by the description of the computational setup and by showing the results, in Sec.~\ref{sec:results}, for a two-dimensional channel and a FCC packing. These will be also compared with other, widely used, traditional definitions of deposition efficiency.

\section{Microscopic equations}\label{sec:micro}
At the pore-scale we assume a simple physical model for non-inertial massless colloidal particles, described by a linear advection-diffusion-reaction equation:
\begin{equation}\label{eq:pore}
\frac{\partial \c}{\partial t} + \nabla \cdot \left(\ub \c\right) - \div \left(\diff \nabla \c \right) = -k_\textrm{b} \c \quad\mbox{on}\;\Omega_\ell\subset\Omega\subset\mathbb{R}^3 \, ,
\end{equation}
where $\Omega_{\ell}$ is the fluid volume, $\Omega$ is the total volume (the sum of the fluid and solid volumes), $\c$ is the scalar concentration (mol m$^{-3}$), $\ub$ is the pore-scale velocity (m s$^{-1}$), $\diff$ is the molecular diffusion\footnote{The working assumption here is that any hydrodynamical interactions (i.e.: hydrodynamic retardation) between the particles and the surface of the solid grains are neglected, employing the Smoluchowski-Levich hypothesis for nanoscopic colloidal particles and making the molecular diffusion coefficient present in the pore-scale advection-diffusion equation $\diff$ a constant, obtained for example, via the Stokes-Einstein relation for diffusion of spheres in liquids.} (m$^2$ s$^{-1}$) and $k_\textrm{b}$ (s$^{-1}$) is the (bulk) reaction coefficient; a detailed sketch-up of the geometrical domain considered can be found in Fig.~\ref{fig:domain}.
At the solid boundaries we assume a generic linear mixed boundary conditions of the type: 
\begin{equation}\label{eq:pore_bc}
\diff \frac{\partial \c}{\partial n}=k_{0}\frac{\alpha}{\alpha-1}\c\quad \textrm{on} \, \Gamma \, , \end{equation}
where $n$ is the surface normal, $\alpha$ is the deposition/attachment efficiency, $\Gamma$ is the porous matrix surface area (m$^2$) and $k_0$ is the surface reaction coefficient ($1/s$).

More details about the equivalency of this Eulerian formulation with Lagrangian formulations are reported in the appendix.
Depending on the type of upscaling, different boundary conditions will be set on the external boundary.
When $\alpha=1$, Eq.(\ref{eq:pore_bc}) is the equivalent of setting homogeneous Dirichlet conditions $c=0$ on the solid grains. Otherwise the term $K=k_{0}\frac{\alpha}{\alpha-1}$ can be interpreted as a reaction flux at the solid surface.

\section{Macroscopic equations}\label{sec:macro}

A possible choice for the averaging operator is the volume average (\citep{whitaker1998method} pg. 9), defined as $\frac{1}{|\Omega|} \int_{\Omega_\ell} \cdot \textrm{d}v = \spatavg{ \cdot}$, where $|\Omega|$ is the total volume (m$^3$) and $|\Omega_{\ell}|$ is the liquid volume (m$^3$). The porosity $\varepsilon$ can be defined as $\varepsilon=\spatavg{1}$.
Applying this operator to Eq.(\ref{eq:pore}) we obtain:
\begin{equation}\label{eq:macro1}
\frac{\partial \langle \c \rangle}{\partial t} + \frac{1}{|\Omega|} \int_{\partial\Omega_{\ell\ell}} \c\ \ub \cdot \mathbf{n}\ \textrm{d}s - \frac{1}{|\Omega|} \int_{\partial \Omega_{\ell\ell}} \diff \nabla \c \cdot \mathbf{n} \textrm{d}s = -\spatavg{ k_\textrm{b} c} - \frac{I}{|\Omega|} \, ,
\end{equation}
having used the Gauss theorem\footnote{$\partial \Omega_\ell = \Gamma \cup \partial \Omega_{\ell\ell}$}, and defining the flux on the solid grains as
$$I=\int_{\Gamma}c\ \ub\cdot\nb-\diff\ \nabla \c \cdot \mathbf{n}\dif s=
\int_{\Gamma}K\c\dif s
$$
where we considered fluid velocity $\ub = 0$ on the grains surfaces.
\begin{figure}\label{fig:domain}
\centering
\begin{tikzpicture}[use Hobby shortcut,closed=true,scale=1.5]

\draw[fill=gray] (0.3,1.3)..(0.7,1.2)..(1.45,0.5)..(1.5,0.35)..(1.5,-0.1)..(1.3,-0.5)..(0.95,-0.9).. (0.6,-1.1)..(0,-1.25).. (-0.5,-1.2)..(-0.8,-1.05)..(-0.9,-1).. (-1.3,-0.6)..(-1.45,-0.3).. (-1.5,0.1)..(-1.4,0.6)..(-1.35,0.72)..(-1.3,0.8)..(-1,1.02).. (-0.7,1.15)..(-0.5,1.225)..(-0.2,1.3)..(0,1.33)..(0.3,1.3);

\draw[fill=white, postaction={pattern=north east lines wide}] (0,0).. (-0.2,0).. (-0.4,0.2) .. (-0.2,0.7).. (-0.1,0.6).. (0.2,0.5).. (0.3,0.4).. (0.3,0.2).. (0,0);
\draw[line width=0.5mm, red] (0,0).. (-0.2,0).. (-0.4,0.2) .. (-0.2,0.7).. (-0.1,0.6).. (0.2,0.5).. (0.3,0.4).. (0.3,0.2).. (0,0);

\draw[fill=white, postaction={pattern=north east lines wide}] (-0.2,-0.2).. (-0.2,-0.4).. (-0.3,-0.5) .. (-0.55,-0.5).. (-0.6,-0.4).. (-0.6,-0.2).. (-0.4,-0.1).. (-0.2,-0.2);
\draw[line width=0.5mm, red] (-0.2,-0.2).. (-0.2,-0.4).. (-0.3,-0.5) .. (-0.55,-0.5).. (-0.6,-0.4).. (-0.6,-0.2).. (-0.4,-0.1).. (-0.2,-0.2);

\draw[fill=white, postaction={pattern=north east lines wide}] (0.2,-0.25)..(0.4,-0.3)..(0.6,-0.5)..(0.4,-0.7)..(0,-0.5)..(0.2,-0.25);   
\draw[line width=0.5mm, red] (0.2,-0.25)..(0.4,-0.3)..(0.6,-0.5)..(0.4,-0.7)..(0,-0.5)..(0.2,-0.25);

\draw[fill=white, postaction={pattern=north east lines wide}] (0.7,-0.25)..(0.5,-0.1)..(0.7,0.35)..(1,0.3)..(1.2,0.1)..(0.7,-0.25);
\draw[line width=0.5mm, red] (0.7,-0.25)..(0.5,-0.1)..(0.7,0.35)..(1,0.3)..(1.2,0.1)..(0.7,-0.25);

\draw[fill=white, postaction={pattern=north east lines wide}] (0.55,0.6)..(0.5,0.8)..(0.8,1)..(1,0.7)..(0.8,0.5)..(0.55,0.6);
\draw[line width=0.5mm, red] (0.55,0.6)..(0.5,0.8)..(0.8,1)..(1,0.7)..(0.8,0.5)..(0.55,0.6);

\draw[fill=white, postaction={pattern=north east lines wide}] (-0.2,1.3)..(-0.2,1)..(-0.1,0.9)..(0.2,0.9)..(0.4,1.1)..(0.3,1.3)..(-0.2,1.3);
\draw[line width=0.5mm, red] (-0.2,1.3)..(-0.2,1)..(-0.1,0.9)..(0.2,0.9)..(0.4,1.1)..(0.3,1.3);

\draw[fill=white, postaction={pattern=north east lines wide}] (1.3,0.8)..(1.2,0.7)..(1.2,0.5)..(1.3,0.4)..(1.5,0.35)..(1.3,0.8);
\draw[line width=0.5mm, red] (1.3,0.8)..(1.2,0.7)..(1.2,0.5)..(1.3,0.4)..(1.5,0.35);

\draw[fill=white, postaction={pattern=north east lines wide}] (1.3,-0.5)..(1.1,-0.4)..(1,-0.5)..(0.55,-0.9)..(0.5,-1)..(0.6,-1.1)..(1.3,-0.5);
\draw[line width=0.5mm, red] (1.3,-0.5)..(1.1,-0.4)..(1,-0.5)..(0.55,-0.9)..(0.5,-1)..(0.6,-1.1);

\draw[fill=white, postaction={pattern=north east lines wide}] (0.3,-1.2)..(0.15,-0.9)..(-0.1,-0.85)..(-0.3,-0.9)..(-0.4,-1)..(-0.5,-1.2)..(0.3,-1.2);
\draw[line width=0.5mm, red] (0.3,-1.2)..(0.15,-0.9)..(-0.1,-0.85)..(-0.3,-0.9)..(-0.4,-1)..(-0.5,-1.2);

\draw[fill=white, postaction={pattern=north east lines wide}] (-0.8,-1.05)..(-0.7,-1)..(-0.7,-0.6)..(-0.9,-0.4)..(-1.3,-0.6)..(-1.2,-1)..(-0.8,-1.05);
\draw[line width=0.5mm, red] (-0.8,-1.05)..(-0.7,-1)..(-0.7,-0.6)..(-0.9,-0.4)..(-1.3,-0.6);

\draw[fill=white, postaction={pattern=north east lines wide}] (-1.4,-0.45)..(-1.3,-0.3)..(-1.3,-0.1)..(-1.5,0.1)..(-1.35,-0.4);
\draw[line width=0.5mm, red] (-1.4,-0.45)..(-1.3,-0.3)..(-1.3,-0.1)..(-1.5,0.1);

\draw[fill=white, postaction={pattern=north east lines wide}] (-1,0)..(-0.6,0.2)..(-0.6,0.3)..(-0.9,0.45)..(-1.2,0.4)..(-1.2,0.1)..(-1,0);
\draw[line width=0.5mm, red] (-1,0)..(-0.6,0.2)..(-0.6,0.3)..(-0.9,0.45)..(-1.2,0.4)..(-1.2,0.1)..(-1,0);

\draw[fill=white, postaction={pattern=north east lines wide}] (-1.3,0.8)..(-1,0.65)..(-0.9,0.7)..(-0.6,0.7)..(-0.6,1.1)..(-0.9,1.2)..(-1.3,0.8);
\draw[line width=0.5mm, red] (-1.3,0.8)..(-1,0.65)..(-0.9,0.7)..(-0.6,0.7)..(-0.575,0.72)..(-0.67,1.16);

\draw[line width=0.5mm, blue] (0.3,1.3)..(0.7,1.2)..(1.1,1)..(1.3,0.8);
\draw[line width=0.5mm, green] (1.3,0.8)..(1.45,0.5)..(1.5,0.35);
\draw[line width=0.5mm, blue] (1.5,0.35)..(1.5,-0.1)..(1.3,-0.5);
\draw[line width=0.5mm, green] (1.3,-0.5)..(0.95,-0.9)..(0.6,-1.1);
\draw[line width=0.5mm, blue] (0.6,-1.1)..(0.3,-1.2);
\draw[line width=0.5mm, green] (0.3,-1.2)..(0,-1.25)..(-0.5,-1.2);
\draw[line width=0.5mm, blue] (-0.5,-1.2)..(-0.8,-1.05);
\draw[line width=0.5mm, green] (-0.8,-1.05)..(-0.9,-1)..(-1.3,-0.6);
\draw[line width=0.5mm, blue] (-1.3,-0.6)..(-1.4,-0.45);
\draw[line width=0.5mm, green] (-1.4,-0.45)..(-1.45,-0.3)..(-1.5,0.1);
\draw[line width=0.5mm, blue] (-1.5,0.1)..(-1.4,0.6)..(-1.35,0.72)..(-1.3,0.8);
\draw[line width=0.5mm, green] (-1.3,0.8)..(-1,1.02)..(-0.7,1.15);
\draw[line width=0.5mm, blue] (-0.7,1.15)..(-0.5,1.225)..(-0.2,1.3);
\draw[line width=0.5mm, green] (-0.2,1.3)..(0,1.33)..(0.3,1.3);
\end{tikzpicture}
\caption{Schematic representation of the reference volume: $\Omega_\ell$ gray, $\Gamma$ red, $\partial \Omega_{\ell\ell}$ blue and $\partial \Omega_{ss}$ green. }
\end{figure}
Assuming a box with periodic boundary conditions on $y-$ and $z-$directions, it is possible to identify the sum of the second and third terms on the LHS of Eq.(\ref{eq:macro1}) with the total flux at the external boundary of the domain, and we can write:
\begin{equation}
\dpd{ \spatavg{\c}}{ t} + \frac{( F_\textrm{tot}^\textrm{out}-F_\textrm{tot}^\textrm{in} )}{|\Omega|}  = -\spatavg{ k_\textrm{b} c} - \frac{I}{|\Omega|}  \, .
\end{equation}
At the steady state, under the hypothesis that the overall source term can be written as the product of a macroscopic effective reaction rate and the averaged concentration, we have:
\begin{equation} \frac{( F_\textrm{tot}^\textrm{out}-F_\textrm{tot}^\textrm{in} )}{|\Omega|} = -\spatavg{k_\textrm{b} c} - \frac{I}{|\Omega|} = - K_\textrm{eff}\spatavg{c} \, ,
\end{equation}
and therefore:
\begin{equation}
K_\textrm{eff}=\frac{F_\textrm{tot}^\textrm{in}-F_\textrm{tot}^\textrm{out}}{\spatavg{c} |\Omega|} \, .
\end{equation}

Discarding the bulk reaction, i.e., $k_\textrm{b}=0$, $K_\textrm{eff}$ represents the effective deposition rate. It is important to notice that this quantity is fully computable from pore-scale simulations, simply looking at inlet-outlet fluxes and averaged volume concentration.
Our macroscopic model equation can thus be written as:
\be{macro}
\frac{\partial \spatavg{c}}{\partial t} + \nabla \cdot \spatavg{\ub \c} - \div \spatavg{\diff \nabla \c} = -K_\textrm{eff}\spatavg{c} \quad\mbox{on}\;{\Omega}\subset\mathbb{R}^3 \, ,
\ee
where we have kept the unclosed averages for the advection and diffusion since, as we will see later, they might affect the effective reaction closure. For the moment, we can assume that these terms can be closed with the standard eddy-dispersion hypothesis, defining an equivalent hydrodynamical dispersion coefficient $\disp$, and with an effective advection based on the Darcy velocity $\Ub=\spatavg{\ub}$.
An alternative is to consider a macroscopic equation, with the same form, for the flux-averaged concentration. This, in principle equivalent under the assumed hypotheses of linearity and homogeneity, can give rise to different approximations when applied to pore-scale systems.

The assumptions of linear effective reaction, Fickian dispersion, and Darcyan velocity are only verified in the (long-time, large averaging volume, slow reaction) limit.
In particular, when the reaction constant $K$ is large \citep{battiato2011applicability}, the effective velocity, dispersion and reaction rate are not decoupled anymore and one should consider regimes where, for example, the effective scalar velocity and dispersion are significantly different from those in the non-reactive case. In this work we will neglect this combined effect and we will focus solely on the effective reaction (deposition, heat transfer) rate. In our future work, after factoring out the effective reaction from the pore-scale equations, we will study its effect on other effective properties, as it has been shown rigorously with homogenisation theory \citep{allaire2007homogenization,allaire2010two}.

\eq{macro}, with the above assumptions and in a simplified one-dimensional dimensionless macroscopic scenario, can be rewritten in a closed form as a macroscopic advection-diffusion-reaction transport equation for $C=(\spatavg{c}/C_{\infty})$:
\be{macro_adim}
\frac{\partial \C}{\partial t_\textrm{diff}} + \varepsilon\PE \dpd{C}{X} - \frac{\disp}{\diff}\dpd[2]{C}{X} = -\DAD\C \quad\mbox{on}\;\Omega\subset\mathbb{R} \ ,
\ee
where $X$ represents the (dimensionless) macroscopic space variable and we have defined $t_\textrm{diff}=\frac{t \diff}{L^{2}}$, being $L$ a characteristic length. Alternatively, defining a dimensionless time based on advection as $t_\textrm{adv}=\frac{T U}{L}$,
\begin{equation}\label{eq:macro_adim2}
\frac{\partial \C}{\partial t_\textrm{adv}} + \dpd{C}{X} - \frac{\disp}{\diff} \frac{1}{\varepsilon\PE} \dpd[2]{C}{X} = -\DA\C \quad\mbox{on}\;{\Omega}\subset\mathbb{R} \ ,
\end{equation}
where $\frac{\disp}{\diff}\frac{1}{\varepsilon\PE}$ is the inverse of the macroscopic P\'eclet number\footnote{For P\'eclet numbers sufficiently high, this becomes a constant, due to the well known linear relation between hydrodynamical dispersion and the velocity, i.e., $\frac{\disp}{\diff} \propto\PE$, for $\PE>1$. }
and the Dahmk\"oler numbers are defined as
\be{DAD}
\DAD=\frac{K_\textrm{eff}\ L^{2}}{\diff}\ ,\qquad \DA=\frac{K_\textrm{eff}\ L}{\U}\ ,
\ee
The similarity between the Sherwood/Nusselt number \eq{Sherwood} and $\DAD$ is evident, the only difference relying on the macroscopic interpretation of microscopic mass transfer at the pore boundaries as an effective volumetric reaction rate.
Despite the striking simplicity of the derivation above, it highlights several aspects commonly neglected in the colloid filtration theory such as the relation between dispersion and reaction.


\subsection{Colloid filtration theory}
To compare the above definition of $K_\textrm{eff}$ with the classical approaches of colloid filtration theory, neglecting the porosity and geometry-dependent constants, we can define three differently defined Dahmk\"oler numbers, based on the concept of $\eta$ (as described in Section 1.2):
\begin{equation}\label{eq:etaAD}
\eta^{A+D}=
\frac{F_\textrm{tot}^\textrm{in}-F_\textrm{tot}^\textrm{out}}{F_\textrm{tot}^\textrm{in}}=1-\frac{F_\textrm{tot}^\textrm{out}}{F_\textrm{tot}^\textrm{in}}=\DA\frac{L}{d_{g}}
\end{equation}

\begin{equation}
\eta^A=\frac{F_\textrm{tot}^\textrm{in}-F_\textrm{tot}^\textrm{out}}{F_\textrm{adv}^\textrm{in}}
=\DA\frac{L}{d_{g}}
\end{equation}

\begin{equation}\label{eq:eta_F}
\tilde{\eta}
=-\log\left(\frac{F_\textrm{tot}^\textrm{out}}{F_\textrm{tot}^\textrm{in}}\right)=-\log(1-\eta^{A+D})
=\DA\frac{L}{d_{g}}
\end{equation}
%
where $F_\textrm{tot}$ is the total flux and $F_\textrm{adv}$ is the advective flux.
The last definition \eq{eta_F} comes from the correct interpretation of $\eta$ as a probability.
Moreover, it can be seen that the first definition \eq{etaAD} is easily comparable to $K_\textrm{eff}$ as follows:
\begin{equation}
\eta^{A+D}=\frac{K_\textrm{eff}\langle c \rangle \arrowvert \Omega\arrowvert}{F_\textrm{tot}^\textrm{in}} = K_\textrm{eff} \langle\tau\rangle
\end{equation}
where the difference between $K_\textrm{eff}$ and $\eta$ can be interpreted as a mean residence time $\spatavg{\tau}$.


\section{Numerical upscaling}\label{sec:numerical}
To test and understand the appropriateness  and consistency of the different definitions of upscaled parameters, we have performed microscopic simulations in simple periodic geometries and computed the effective deposition rates for different value of $\PE$ and the reaction rate at the surface, $K$.

\subsection{Computational setup}
Two different geometries were chosen: a bi-dimensional channel (CH) and a face-centered periodic sphere packing (FCC).
The purpose of the first is for a validation of the modelling and numerical approaches, since analytical results exist for heat/mass transfer coefficients in such a simple domain~\citep{incropera1985fundamentals,Bird1960}.
Although the dimensionless version of the Stokes and continuity equation and of \eq{pore} have been solved, the simulations are equivalent (just to give a realistic example) to a channel $2\times$\Die{-3}~m wide and \Die{-2}~m long and to FCC spheres with a diameter of $2\times$\Die{-4} m with a porosity equal to 0.4. In this way, for the FCC geometry the length of the periodic cell is equal to $3.03\times$\Die{-4} m.  A wall-adapted mesh, equivalent to a $100^{3}$ resolution has been used. Our previous studies \citep{IBMTS2014,P003,P006} shown that this resolution give a reasonably low ($<1\%$) error both in average flow, dispersion and reaction rate.

Simulations were performed with the \textsf{OpenFOAM 4.x} library \citep{oF}.
First, fluid flow simulations were performed with the \textsf{simpleFoam} solver.
Full periodicity has been set at the boundaries with a uniform forcing term to represent the role of the constant pressure gradient condition, to obtain the desired mean velocity; simulations were performed under creeping flow conditions, compatibly with groundwater and filtration applications.
For transport simulations, a parametric sweeping on the values of the $\PE$ number was performed, as $\PE$ ranges from \Die{-2} to \Die{3}; the values of the diffusion coefficient $\mathcal{D}_0$ are chosen according to the desired \PE.
A modified and extended version of \textsf{scalarTransportFoam} solver has been used; particular attention has been devoted to the implementation of the boundary conditions, as described in the following paragraph.


\subsubsection{Scalar boundary conditions}
A novel \textit{pseudo-periodic} approach that allows to quickly obtain a converged asymptotic solution has been implemented.
While the  conditions on the lateral boundaries are periodic, the same is not possible at the inlet and outlet boundaries due to the non-conservative nature of the transport equation.
The concentration profile at the inlet has therefore been set to be proportional to the outlet profile, and rescaled to have a fixed unitary mean value.
The linearity of the equation, in fact, allows us to define the solution up to a multiplicative constant.
Thanks to this setup, the self-similar asymptotic concentration profiles are found and a stationary solution of the equations exists and is obtained either with iterations in time or with non-linear iterations of the time-independent equation.
In both cases, the solution is obtained after approximately one advective time scale, i.e., the time needed for the information to travel from the inlet to the outlet.
This is particularly effective for very large P\'eclet numbers, that would have otherwise needed very long computational domains and long times to observe the asymptotic profiles.
At the solid boundaries, the mixed (Robin) condition \eq{pore_bc} is applied with a reaction constant $k_0=1$ (m s$^{-1}$). As described in the appendix, both boundary conditions have been implemented in \textsf{OpenFOAM} by means of modifications, respectively, of the \textsf{mixed} and \textsf{mappedBC} boundary conditions.

\section{Results}\label{sec:results}

\begin{figure}[t!]
\begin{center}
\makebox[\textwidth][c]{
\includegraphics[width=.8\textwidth]{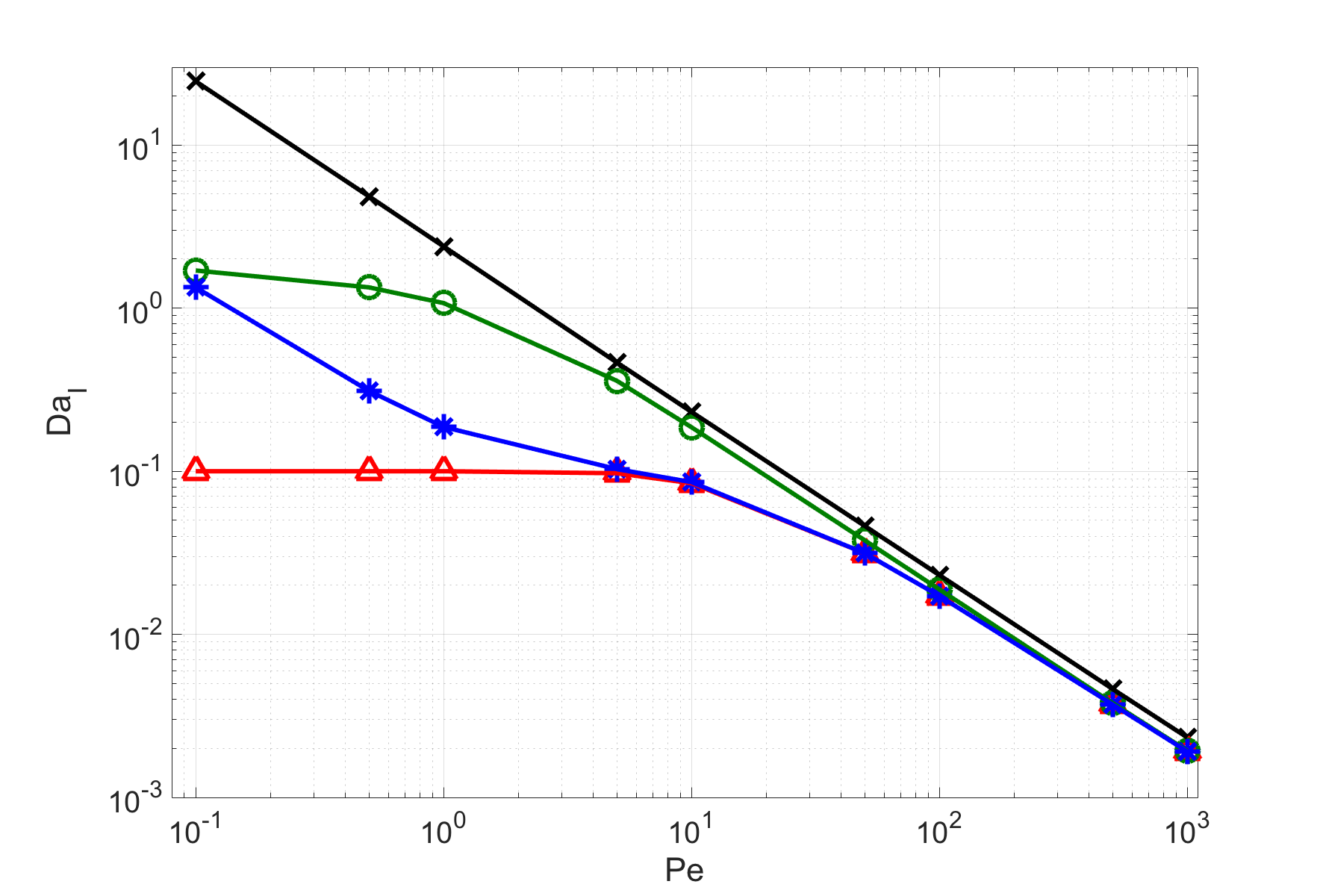}
}
\caption{Two-dimensional channel. Effective reaction rate $\DA$ against $\PE$. $K_\textrm{eff}$ (black line with crosses), $\tilde{\eta}$ (green line with circles), $\eta^{A}$ (red line with triangles), $\eta^{A+D}$ (blue line with stars).}
\label{fig:channel}
\end{center}
\end{figure}

As a first consistency and verification step, a simulation of a two-dimensional channel flow has been performed with $\alpha=1$ (ideal deposition, homogeneous Dirichlet condition).
In this case, classical and well-known analytical results exist for heat/mass transfer phenomena.
In particular, it can be shown~\citep{incropera1985fundamentals} that the effective Sherwood/Nusselt number for a fully developed flow and concentration profile, is constant, independent of $\PE$.
This is verified in~\fig{channel}, showing results for a channel with a $10\times 1$ length-to-height ratio, with the proposed pseudo-periodic computational setup.
Here, the black line with crosses represents $\DA$, computed with $K_\textrm{eff}$, according to \eq{DAD}, for different values of the P\'eclet number. It is important to notice that, without the pseudo-periodic setup, the length of the channel should have been extremely (and unfeasibly) long to be able to reach the stationary profile.
In accordance with the definitions \eq{Sherwood} and \eq{DAD}, a linear inverse relation between $\DA$ and $\PE$ is found (therefore $\DAD$ being constant). The other curves represents the other upscaling approaches based on $\eta$, namely $\tilde{\eta}$ (green line with circles), $\eta^{A}$ (red line with triangles), and $\eta^{A+D}$ (blue line with stars). All of them, as expected, are diverging from the correct volume averaged reaction rate for low P\'eclet number, with the last one converging to $0.1$, since they have been normalised by the channel length.

\begin{figure}[t!]
\begin{center}
\makebox[\textwidth][c]{
\includegraphics[width=.8\textwidth]{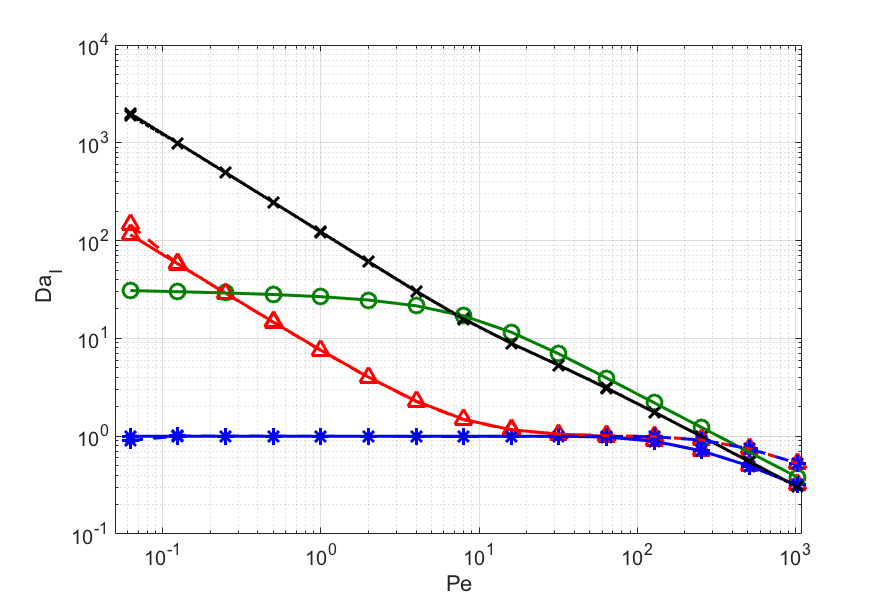}
}
\caption{FCC packing. Effective reaction rate $\DA$ against $\PE$. 1 module (continuous lines), 2 modules (dashed lines). Upscaling based on $K_\textrm{eff}$ (black line with crosses), $\tilde{\eta}$ (green line with circles), $\eta^{A}$ (red line with triangles), $\eta^{A+D}$ (blue line with stars).}
\label{fig:fcc}
\end{center}
\end{figure}

\begin{figure}[t!]
\begin{center}
\makebox[\textwidth][c]{
\includegraphics[width=.8\textwidth]{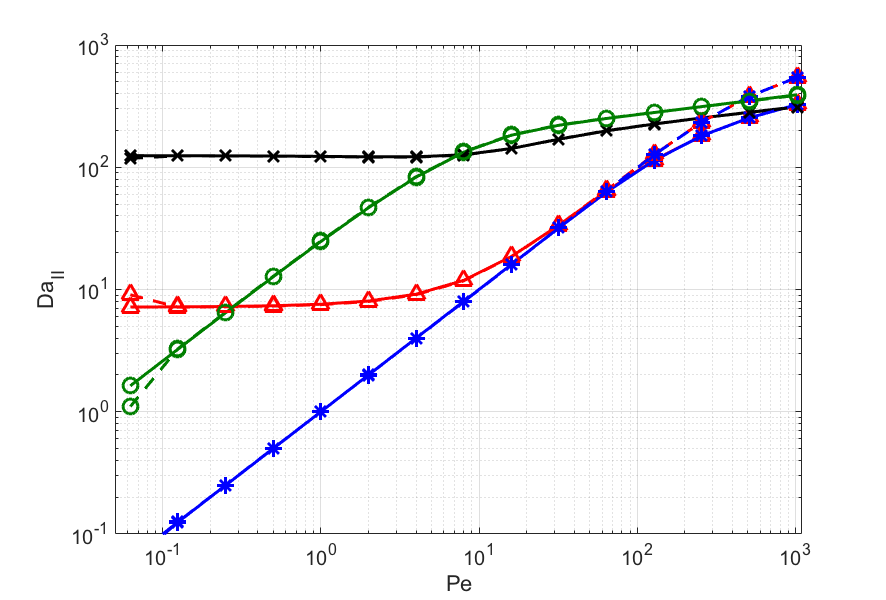}
}
\caption{FCC packing. Effective reaction rate $\DAD$ against $\PE$. 1 module (continuous lines), 2 modules (dashed lines). Upscaling based on $K_\textrm{eff}$ (black line with crosses), $\tilde{\eta}$ (green line with circles), $\eta^{A}$ (red line with triangles), $\eta^{A+D}$ (blue line with stars). }
\label{fig:fcc2}
\end{center}
\end{figure}

Using the same notation, the simulations performed also on a periodic FCC arrangement are reported in \fig{fcc}. Here, to verify that the upscaling concepts are scale-independent, also simulations with two adjacent cells are performed (dashed lines). As it can be seen, not only the upscaling based on $K_\textrm{eff}$ shows a certain physical consistency at all P\'eclet numbers ($\PE^{-1}$ rate followed, for $\PE>10$, by a $\PE^{-0.85}$ rate), but it is also completely scale-independent. On the other hand, the approaches based on $\eta$, despite the normalisation by the total length, are scale-dependent ($\eta^{A}$ and $\eta^{A+D}$, for $\PE>100$) and do not show a consistent (and simple) behaviour for $\PE<1$. This also confirms how, the interpretation of $\eta$ as an efficiency ($0<\eta<1$), and all its extensions cannot give consistent effective reaction rates that, for vanishing $\PE$ has to tend to infinity.

When the reaction rate is considered with the diffusive scaling (e.g., $\DAD$ or, equivalently, $\SH$ or $\NU$), in \fig{fcc2}, we can notice how, compared to the cases of walls parallel to the flow (channels and pipes), the presence of the complex geometry create a transition, for $\PE>10$ towards a non-constant regime, that will be further detailed below.

\subsection{Effect of attachment efficiency $\alpha$}

\begin{figure}[t!]
\begin{center}
\makebox[\textwidth][c]{
\includegraphics[width=.75\textwidth]{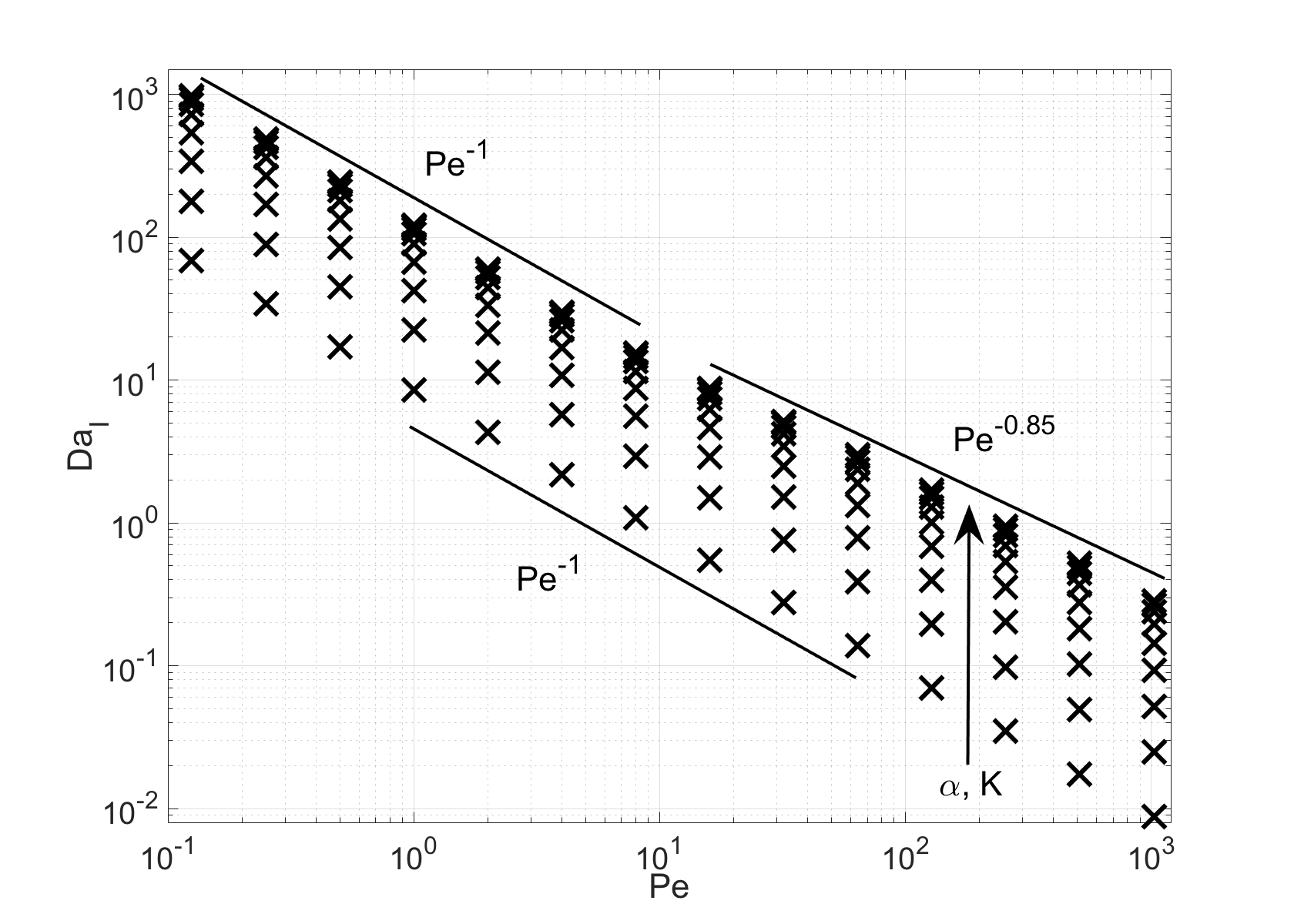}
}
\caption{FCC packing. $\DA$ against $\PE$.}
\label{fig:fcc_alpha}
\end{center}
\end{figure}

\begin{figure}[t!]
\begin{center}
\makebox[\textwidth][c]{
\includegraphics[width=.75\textwidth]{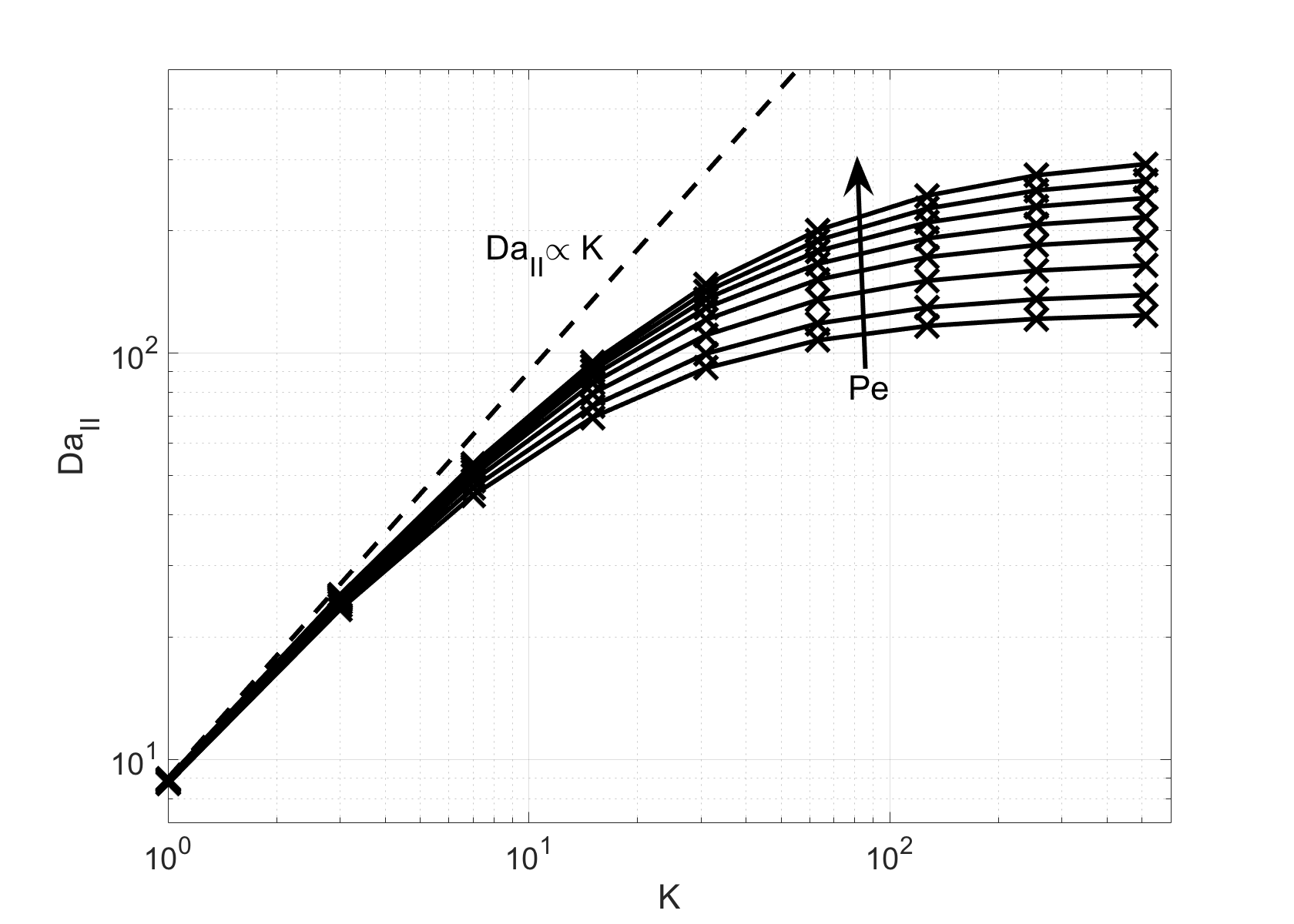}
}
\caption{FCC packing. $\DAD$ against the surface reaction rate $K$.}
\label{fig:fcc_alpha2}
\end{center}
\end{figure}

When considering partial absorption ($\alpha<1$), the qualitative behaviour of the effective reaction rate, as expressed by $\DA$ and $\DAD$, defined by $K_\textrm{eff}$, already observed for $\alpha=1$, persists. In particular, as it is shown in \fig{fcc_alpha}, where $\DA$ is plotted against $\PE$ for all values of the surface reaction rate $K$, the regime transition appears at the same P\'eclet number ($\approx 10$). However, for finite $K$, the second regime $\PE^{-0.85}$ does not persist and seems to fall back into the $\PE^{-1}$ rate for sufficiently high $\PE$, where this second regime transition happens for $\PE\approx K$, i.e., the microscopic surface Dahmk\"oler number matches the microscopic P\'eclet number. It is important to notice here that, in our results, $K$ takes the role of a surface Dahmk\"oler number, considering $\diff=1$ and a unitary length scale $L=1$. In a more general situation, the discussion of this section should be done considering the adimensional number $\frac{K L}{\diff}$ instead.

In \fig{fcc_alpha2}, the dependence of $\DAD$ on the surface reaction rate $K$ is plotted for different $\PE$. As expected, for slow reactions, $K\lesssim 1$, the overall effective reaction rate is linear with $K$, while, for $K\to\infty$ converges to the (possibly $\PE$-dependent) limit we have studied in the results above with $\alpha=1$. These results suggest the following conclusions:
\be{rates}
\DA\propto\begin{cases}
\PE^{-1}\qquad &\mbox{for}\quad \PE\lesssim10 \\
\PE^{-0.85}\qquad &\mbox{for}\quad 10\lesssim\PE\lesssim \frac{K L}{\diff} \\
\PE^{-1}\qquad &\mbox{for}\quad \PE\gtrsim \frac{K L}{\diff}
\end{cases}
\ee
where we remind that $K$ represents the reaction rate at the surface and the exponent $0.85$ is characteristic of the FCC geometry.


\section{Conclusions}\label{sec:conclusions}
In this work we propose a volume averaging approach to derive a consistent definition of deposition efficiency, written as a volumetric effective reaction rate $K_\textrm{eff}$, and we reconcile the classical studies of heat and mass transfer to the classical theory of clean bed filtration and its recent computational studies. In particular we discuss current limitations and some inconsistencies of the standard approach based on the definition of a deposition efficiency, $\eta$. This parameter, usually defined as a percentage of deposited particles on a discrete finite geometry, cannot be used directly at the macroscopic scale for an arbitrary P\'eclet number and cannot be consistently defined on arbitrary pore-scale domains. We also propose a fast and efficient computational setup to compute the asymptotic upscaled reaction parameters effectively, including the effect of partial absorbing boundaries with an arbitrary attachment probability/efficiency $\alpha$. This is another parameter that is often simply multiplied to the macroscopic reaction rate. We show here that a boundary reaction rate $K\propto\frac{\alpha}{\alpha-1}$ can be defined and, only in the limit of $\alpha\to 0$, the macroscopic deposition rate  behaves linearly, i.e, $K_\textrm{eff}\propto K\propto \alpha$.   Our future studies will involve a further investigation of the large P\'eclet and large Dahmk\"oler number limit, comparison with homogenisation theory, and an analysis of additional DLVO physical models, and non-linear extensions and isotherms for the solid boundary conditions.

\section*{Acknowledgements}
Financial support for this work has been provided by our rich and generous aunts from Pinerolo, whom we gratefully acknowledge.

\section*{References}
\bibliographystyle{chicago}
\bibliography{depBib}

\clearpage

\appendix

\section{Drift-diffusion Lagrangian equivalent}\label{sec:lagrangian}

It is well known that the Ito SDE
\begin{equation}\label{eq:sde_pos}
\dif \Xb= \ub(\Xb)\dif t + \sqrt{2\diff}\dif \wienerb
\end{equation}
with $\Xb(t=0)=\Xb_{0}$, where $\wienerb$ is a n-dimensional Brownian motion, $\diff$ a diffusion constant, and $\u$ is a space-dependent constant velocity field, is equivalent, in an unbounded domain, to the Advection-Diffusion (Fokker-Planck) equation
\be{fp_pos}
\dpd{\f}{t}=-\divx\del{\ub(\xb)\f}+\divx\del{\grad{\diff\f}}
\ee
with $\f(\xb;t=0)=\delta(\xb-\Xb_{0})$, and $\f=\f(\xb;t)$ defined as
\be{fp_def}
\f=\ensavg{\delta(\Xb-\xb)}
\ee
and the the operator $\ensavg{\cdot}$ indicates the average with respect to the Brownian motion (stochastic average).
However, in the presence of boundaries and inhomogeneous or non-linear boundary conditions, the connection between the two formulations is not straightforward and can lead to several conceptual and numerical errors.
In particular, while the cases for absorbing (homogeneous Dirichlet condition $\f=0$) and reflecting (homogeneous Neumann condition $\partial_{n}\f=0$) boundaries can be simply related to simple discrete Lagrangian rules (respectively, the elimination of the particles when crosses the boundary or the symmetric reflection), the partially reflecting case lead to several conceptual problems, studied in the probability literature as the \textit{partially reflected Brownian motion} and the similar \textit{Sticky Brownian motion}\citep{feller1954diffusion}.
For example, it is well known that a Brownian particles starting at a point crosses infinitely many times that point in a finite time interval. This means that, if the particle, every time it hits the wall, is absorbed with probability $\alpha$ or reflected with probability $1-\alpha$, this will result in the particle being always absorbed. This does not happen numerically when a Brownian motion is discretised with finite time intervals\citep{singer2008partially}, creating a dangerous dependence of the results on the time discretisation and a clear inconsistency between the Lagrangian and Eulerian pictures. A simple alternative is to define a fixed distance from the boundary, $a$, to which the particle jumps every time it hits the wall and has been reflected. \citet{grebenkov2006partially} shows how to formalise this stochastic process and its equivalence to the Robin condition
\be{mixedbc}
\diff\dpd{\f}{n}=\frac{a\, \alpha}{\alpha-1}\f
\ee 

Despite this analogy, the numerical implementation of Lagrangian deposition scheme is a particularly delicate task since, minor changes in the implementation or time stepping can give quantitatively and qualitatively different equivalent Eulerian formulation\citep{singer2008partially}. We believe, therefore, that the Eulerian approach is preferable and further studies on how to parametrise more complex physical models (such as DLVT theory) in this framework will be carried out.

\section{OpenFOAM implementation}\label{sec:openfoam}
The entire suite of tools used for the simulations reported in this work are released open-source\cite{bitbucket}
\subsection*{Mapped BCs}
The \textsf{mapped} boundary conditions has been modified to preserve the profile up to a normalising constant. This can be done keeping constant the total inward flux or the mean concentration. The first approach, despite being more realistic, it also involves the velocity field and results therefore in being slightly more unstable. A more accurate approach is to map no only the concentration value but also its normal derivative, both scaled.

\subsection*{Mixed BCs}
The condition \eq{pore_bc} can be implemented using the \textsf{codedMixed}  or \textsf{mixed} boundary conditions, i.e. a condition that computes the boundary value as
\be{openfoam_mixed}
\c_{p}=w V + (1-w) \left(\c_{n}+\frac{G}{\Delta}\right)\,,
\ee
where $V$ is the reference value, $G$ is a reference normal gradient, $\c_{p}$ is the boundary value, $\c_{n}$ is the cell centre value, with $\Delta$ the corresponding distance. However, since \eq{pore_bc} can be approximated as
\begin{equation}\label{pore_bc_lin}
\frac{\c_{n}-\c_{p}}{\Delta}=\frac{\alpha}{\alpha-1}\c_{p}\qquad\Rightarrow\qquad \c_{p}=\c_{n}\frac{(\alpha-1)}{\alpha\Delta + \alpha -1}\,,
\end{equation}
it resulted more convenient to develop a new boundary condition, which we simply named \textsf{Robin}, that implements a generic linear BC of the form
\be{robin}
\diff\dpd{\c}{n}=K\c + F
\ee
where, in our case, $K=\frac{\alpha}{\alpha-1}$ and $F=0$.

\end{document}